\documentclass{ieeetran}
\usepackage{amsmath,amssymb,amsfonts}
\usepackage{graphicx}
\usepackage{algorithm}
\usepackage[noend]{algpseudocode}

\def\BibTeX{{\rm B\kern-.05em{\sc i\kern-.025em b}\kern-.08em
    T\kern-.1667em\lower.7ex\hbox{E}\kern-.125emX}}

\begin{document}


\title{Design and Analysis of Polar Codes Based on Piecewise Gaussian Approximation}

\author{R. M. Oliveira and R. C. de Lamare
\thanks{This work was supported by CAPES, CNPq, FAPERJ and FAPESP. The authors are with the Centre for Telecommunications Studies (CETUC), Pontifical Catholic University of Rio de Janeiro (PUC-Rio), Rio de Janeiro-RJ, Brasil. Emails: rbtmota@gmail.com,delamare@cetuc.puc-rio.br}}



\maketitle

\begin{abstract}
In this article, we propose the construction of polar codes based on piecewise Gaussian approximation (PGA) techniques. The PGA is first optimized and then compared to the Gaussian approximation (GA) construction method, showing performance gains for medium blocks and high precision for long blocks, in scenarios with successive cancellation (SC) decoding and additive white gaussian noise (AWGN) channel. Based on the PGA, we develop two approximations based on multi-segmented polynomials that are easy to implement. We present the Approximate PGA (APGA) that is optimized for medium blocks and provides a performance improvement without increasing complexity. Furthermore, we develop the simplified PGA (SPGA) as an alternative to the GA, which is optimized for long blocks and achieves high construction accuracy. Simulation results show that the APGA and SPGA construction methods outperform existing GA and competing approaches for medium and long block codes with notable performance improvement.

\end{abstract}



\maketitle

\section{Introdution}
In 2009, polar codes (PC) were introduced by Ar\i kan \cite{Arikan}. PC are the first channel coding scheme to achieve channel capacity with low encoding and decoding complexity. Due to this and their excellent performance, PC were selected for the fifth generation (5G) \cite{5G} wireless communications standard.

The construction of PC consists of the calculation of the channel reliability and the selection of the locations of the information bits, i.e., the most reliable channel locations will be used for the information bits \cite{Arikan}. When the block length tends to infinity, the most reliable channels are also called noiseless channels. Several construction methods have been proposed in the last decade such as the Bhattacharyya parameter (BP) and Monte Carlo (MC) \cite{Arikan}; density evolution (DE) \cite{Mori1},\cite{Mori2},\cite{Tal1}; Gaussian approximation (GA) of density evolution \cite{Chung},\cite{Trifonov}; and the polarization weight (PW) \cite{RZhang}\cite{Schurch}. The BP approach \cite{Arikan} is a recursive heuristic method that is excellent for the Binary Erasure Channel (BEC) as well for other Binary Discrete Memoryless Channels (B-DMCs). The MC method \cite{Arikan} is an exhaustive search method based on SC. The study in \cite{Vangala} demonstrates the use of the BP and the MC methods in the construction of PC over AWGN channels. The DE approach has been proposed by Mori and Tanaka \cite{Mori1},\cite{Mori2}, which theoretically has the highest accuracy. To obtain the error probability of each subchannel in a simplified form, Tal and Vardy \cite{Tal1} proposed a method to calculate the upper bound and the lower bound of this probability. The polarization weight (PW) \cite{RZhang}\cite{Schurch} is a channel-independent approximation method, which estimates the channel reliability as a function of its index. Polar codes can also be constructed and adapted to a specific decoder, for example, construction of polar codes for List SC decoding \cite{P. Yuan} and Belief Propagation (BP) decoding \cite{M. Qin}\cite{S. Sun}. In \cite{A. Elkelesh} the authors propose a genetic algorithm framework that jointly optimizes the PC construction and rate with a specific decoder. The its construction obtained for BP decoder has performance compared to SC decoder with list. Deep learning-based polar code design proposed in \cite{M. Ebada} allows one to optimize codes for any finite block length, decoder and channel type considering its noise statistics. They obtained a polar code construction for BP decoder over Rayleigh fading channel with superior performance.

Chung \cite{Chung} introduced GA and applied it to Low-Density Parity-Check (LDPC) code construction, whereas other approaches for LDPC code design have employed graph optimization \cite{memd}. The work in \cite{Trifonov} was the first to use it in the construction of polar codes. GA was originally described in integral form, known as exact GA (EGA). Due to the complex integration, EGA has a high computational cost associated with the numerical solution, which increases exponentially with the code length and with the polarization levels. As an alternative, Chung \cite{Chung} proposed the approximate GA (AGA), which is an approximation composed of a two-segment function. We note that this alternative is implemented by transcendental functions, maintaining a high computational complexity. The author in \cite{Jeongseok} also proposed an alternative to numerical integration, approximating GA by a three-segment function. Trifonov also proposed in \cite{Trifonov2} a multi-segment polynomial approximation, without the use of transcendental and inverse functions. The AGA performance for short and medium blocks for the AWGN channel is similar to the Tal and Vardy method, but it fails for long blocks due to the approximation error around zero. Examples of algorithms and performance comparisons can be found in \cite{Vangala} and \cite{Cheng}.

For long blocks, the PC construction with EGA is expensive due to the numerical integration, in addition to the errors associated with the numerical integration solution, and the AGA construction is imprecise due to the approximation error around zero. Furthermore, both require inverse, transcendental functions and are composed of complex recursive functions. Improved approximations with better performance than AGA \cite{Chung} were proposed in \cite{Fang}, \cite{Dai} and \cite{Ochiai}. The work of Fang \cite{Fang} analyzed the first and second derivatives of EGA and developed a simplified multi-segment polynomial approximation without using transcendental functions and without the need to calculate an inverse function. Dai \cite{Dai} introduced the concepts of polarization violation set, the polarization reversal set and a new metric named cumulative-logarithmic error, which results in an algorithm that uses transcendental and inverse functions. Ochiai et al. \cite{Ochiai} analyzed the behavior of EGA in the logarithmic domain and proposed another approximation based on a logarithmic function (transcendental function) and that employs an algebraic expression for the inverse function.

Given the computational complexity of EGA related to the complex integration, the intrinsic imprecision of the AGA associated to the approximation error around zero, and that the approximations proposed by \cite{Fang}, \cite{Dai} and \cite{Ochiai} employ transcendental functions or require function inversion for PC construction, i.e., both the original GA method and the previous approximations have numerical integration and function inversion, we propose an approximation function that replaces the numerical integration and the inverse function with a set of piecewise polynomial functions, resulting in an improved approximation and further computational simplicity. Then, we develop in this work two improved approximations for GA based on PGA. In particular, we develop high-precision approximations using only multi-segment polynomial functions, which replace the need for numerical integration, function inversion and transcendental functions. In \cite{Oliveira5} we reported the preliminary results. Specifically, we have expanded the work in \cite{Oliveira5} by including the application of PGA and extended design techniques to medium and large blocks, with theoretical analysis and extra simulation results of various application scenarios. In particular, we devise a novel strategy for a piecewise approximation method for PC construction, resulting in improved performance for medium block lengths. Similar to the original GA function, PGA is used in integral form. Then, we propose an approximation called Approximate PGA (APGA), through a new criterion of the approximation inspired by a detailed analysis of the behavior of the PGA function. APGA is a simplified alternative multi-segment polynomial approximation, which is computationally more convenient for construction of PC with medium blocks. By drawing inspiration from the analysis, we devise an approximation for long blocks, called Simplified PGA (SPGA), also in the form of a multi-segment polynomial function. The proposed method can be generalized to extremely long or extremely short lengths and is able to adapt to any channel condition. Moreover, we show that the difference in accuracy between the approximation methods can be obtained by the Number of Different Positions (NDP), initially introduced by Kern \cite{Kern}, and we derive an index that measures the general quality of the proposed approximation, called Accumulative Design Error (ADE). The rate-compatible PC design that uses GA in its construction, as reported in \cite{Oliveira1}, \cite{Oliveira2}, \cite{Oliveira3}, \cite{Oliveira4}, can have a significant impact on its performance when considered with the APGA and EPGA constructions.

The main contributions of this article are summarized as:
\begin{itemize}
  \item We propose the PGA construction method for PC;

  \item We develop the APGA construction \cite{Oliveira6} for medium blocks based on the detailed analysis of the behavior of the PGA function, the identification of its key points and the analysis of the statistical distribution of the results generated by its function;

  \item Using the same analysis criterion, we devise a novel approximation for long blocks, also in the multi-segment function form, called SPGA;

  \item We propose the use of ADE as a comparison index between EGA construction methods and SPGA to assess the quality of the approximation for long blocks.

  \item A comparative study in terms of Frame Error Rate (FER) performance analysis between APGA and SPGA with other existing approximation techniques.
\end{itemize}

In addition to the FER, the bit error rate (BER) is also commonly used as a metric of system performance. However, the FER metric also shows the retransmission effort required in a given communication system, that is, these are effectively the frames that will be retransmitted in the event of an error, regardless of the number of bits in error that occurred.

This paper has the following structure. In Section II, we briefly review the fundamentals of PC. In Section III, the Gaussian approximation is briefly presented. In Section IV, we detail the construction of PC by PGA. In Section V, we present the application of PGA to design PC with medium and long blocks. In Section VI, we show and discuss the numerical results. In Section VII we draw the conclusions of this work.

\section{Polar Codes}

In this section, we review the fundamentals of PC, including their basic definitions, encoding and decoding.

\subsection{Basic definitions}

Given a symmetric binary-input discrete memoryless channel (B-DMC) $W:\mathcal{X} \to \mathcal{Y}$, where $\mathcal{X}$  $= \{0,1\}$ and $\mathcal{Y} \in \mathbb{R}$. We define $W(y|x)$ as the channel transition probability, where $x \in \mathcal{X}$ and $y \in \mathcal{Y}$. In order to transmit the information bits, the most reliable channels are chosen. The indices of these channels are then represented by the set $\mathcal{A}$, whose size is defined by $K$. In turn, $\mathcal{A}^\text{c}$ is its complementary set, containing the indices of the least reliable channels that correspond to the sequence of frozen bits. Polar codes can be completely specified by three parameters as designated by PC$(N,K,\mathcal{A}^\text{c})$, where $N$ is the codeblock length, $K$ the length of the information sequence and $\mathcal{A}^\text{c}$ are the indices of frozen bits. In the B-DMC cases the frozen bits are all-zeros. The $K/N$ ratio is called the code rate $R$.

We write $W^N$ to denote the channel corresponding to $N$ uses of $W$; thus, $W^N: \mathcal{X}$$^N$ $\to \mathcal{Y}$$^N$ as
\begin{equation}
  W^N \left(y_1^N|x_1^N \right)=\prod_{i=1}^N W(y_i|x_i).
\label{eq:01}
\end{equation}

The mutual information is defined by \cite{Arikan} for the B-DMC W channel
\begin{equation}
  I(W) = \sum\limits_{y \in Y}\sum\limits_{x \in X}\frac{1}{2}W(y|x)\log\frac{W(y|x)}{\frac{1}{2}W(y|0)+\frac{1}{2}W(y|1)},
\label{eq:02}
\end{equation}
where the base-2 logarithm $0 \leq I(W) \leq 1$ is employed.

After we apply the polarization process \cite{Arikan} to the $N$ independent channels of $W$, we obtain a set of polarized channels $ W_N ^ {(i)}: \mathcal{X} \to \mathcal{Y} \times \mathcal{X}^{\text{i-1}}$, $i = 1,2,\ldots,N$. As defined in \cite{Arikan}, this channel transition probability is given by
\begin{equation}
  W_N^{(i)}\left(y_1^N,u_1^{(i-1)}|u_i\right) = \sum\limits_{u_{i+1}^N \in X^{N-1}}\frac{1}{2^{N-1}}W_N\left(y_1^N|u_1^N\right).
\label{eq:03}
\end{equation}

According to \cite{Arikan}, $N \to \infty$, {$I(W_N^{(i)})$} tends to $0$ or $1$.

\begin{figure}[htb]
\begin{center}
\includegraphics[scale=0.9]{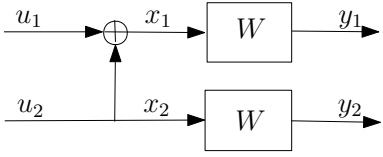}
\caption{The Channel $W_2$}
\end{center}
\label{figura:fig01}
\end{figure}

In Fig. 1 we show the process of creating the channel $W_2$: a recursive step that combines two copies of independent $W$, which have the transition probabilities \cite{Arikan} given by
\begin{equation}
  W_2^{(1)}\left(y_1^2|u_1\right)=\sum_{u_2}\frac{1}{2}W\left(y_1|u_1 \oplus u_2\right)W\left(y_2|u_2\right),
\label{eq:04}
\end{equation}
\begin{equation}
  W_2^{(2)}\left(y_1^2,u_1|u_2 \right)=\frac{1}{2}W\left(y_1|u_1 \oplus u_2\right)W\left(y_2|u_2\right).
\label{eq:05}
\end{equation}

\subsection{Encoding}

The encoding is given by ${\mathbf x}^N_1 = {\mathbf u}^N_1\textbf{G}_N$, where $\textbf{G}_N$ is the transformation matrix, ${\mathbf u}^N_1 \in \{0,1 \}^N$ is the input block and ${\mathbf x}^N_1 \in \{0,1 \}^N$ is the codeword, where ${\mathbf u}^N_1$ is a permutation between ${\mathbf u}_{\mathcal{A}}$ and ${\mathbf u}_{\mathcal{A}^ \text{c}}$, with ${\mathbf u}_{\mathcal{A}}$ containing the bits of information and ${\mathbf u}_{\mathcal{A}^\text{c}}$ the frozen bits. We define $\textbf{G}_N = \textbf{B}_N\textbf{F}^{\otimes n}_2 $, where $\otimes$ denotes the Kronecker product, $\textbf{F}_2 = \footnotesize\left[\begin{array}{cc}
1 & 0 \\
1 & 1 \end{array} \right]$ and $\textbf{B}_N$ is the bit-reversal permutation matrix. A simplification without loss of generalization is the omission of $\textbf{B}_N$.

\subsection{Decoding}

Given the received vector ${\mathbf y}^N_1\in \mathbb{R}^N$, with ${\mathbf y}^N_1$ $=(y_1,\ldots,y_N)$ and ${\mathbf y}^N_1={\mathbf x}^N_1 + {\mathbf n}$, where ${\mathbf n} \in \mathbb{R}^N$ is the noise vector. The objective of the decoder is to obtain estimates of the codeword at the input of the channel ${\mathbf u}^N_1 \in \{0,1 \}^N$  that is given in vector form as $\hat{\mathbf u}^N_1=(\hat{u}_1,\ldots,\hat{u}_N)$. The likelihood ratio (LR) of $u_i$, ${\rm{ LR}}(u_i)=\frac{W^{(i)}_N({\mathbf y}^N_1,\hat{u}_1^{i-1}|0)}{W^{(i)}_N({\mathbf y}^N_1,\hat{u}_1^{i-1}|1)}$ is used with Successive Cancellation (SC) for decoding \cite{Arikan}. Then, the estimated value $\hat{u_i}$ is given by
\begin{equation}
  \hat{u}_i=\begin{cases}
    h_i({\mathbf y}_1^N,\hat{u}_1^{i-1}), & \text{if $i \in \mathcal{A}$},\\
    u_i, & \text{if $i \in \mathcal{A}$$^c$},
  \end{cases}
\label{eq:06}
\end{equation}
where $h_i:{\mathbf y}^\text{N} \times \mathcal{X}$ $^{i-1} \to \mathcal{X}$, $i \in \mathcal{A}$, are decision functions defined as
\begin{equation}
  h_i({\mathbf y}_1^N,\hat{u}_1^{i-1})=\begin{cases}
    0, & \text{if $\frac{W^{(i)}_N({\mathbf y}^N_1,\hat{u}_1^{i-1}|0)}{W^{(i)}_N({\mathbf y}^N_1,\hat{u}_1^{i-1}|1)}\geq 1$},\\
    1, & \text{otherwise},
  \end{cases}
\label{eq:07}
\end{equation}
for ${\mathbf y}_1^N \in \mathcal{Y}^\text{N}$, $\hat{u}_1$ $^{i-1} \in \mathcal{X}$ $^{i-1}$.

The notation $W$ is used for both the channel and its probability
\cite{Arikan}, that is, $W(y|x)$ corresponds to the transition
probability $p(y|x)$ of the channel. Therefore, the term
$W^{(i)}_N({\mathbf y}^N_1,\hat{u}_1^{i-1}|0)$ and
$W^{(i)}_N({\mathbf y}^N_1,\hat{u}_1^{i-1}|1)$ represents the
probability that the bit that was transmitted through the channel be
equal to 0 or 1, respectively. The term ${\mathbf y}_1^N$ is the
entire vector ${\mathbf y}$ and the term $\hat{u}_1^{i-1}$ is the
previous decoded value. Note $u_i$ is input vector, $\hat{u_i}$ is
the decoded codeword and $\hat{u_i}$ is an estimate of $u_i$. We
have that $\hat{u_i}$ is a decoded bit and also an estimate of the
bit $u_i$, as well as $\hat{u}^{i-1}$ is the previously decoded bits
\cite{Arikan}.

We denote $L_N^{(i)}$ as the LR node, $N$ being the index of the row
and $i$ being the stage or the column, following the mapping of the
decoding tree \cite{Arikan}. In Fig. 2 we have a graphic
representation of the SC decoder.

\begin{figure}[htb]
\begin{center}
\includegraphics[scale=1.0]{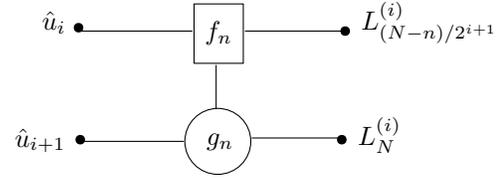}
\caption{Graphic representation of the SC decoder.}
\end{center}
\end{figure}

The values $L_N^{(i)}$ can be obtained recursively using the equations:
\begin{equation}
  L_N^{(i+1)}= \\
   \begin{cases}
    f_n(L_N^{(i)},L^{(i)}_{(N-n/2^{i+1})}), &\text{($f_n$ nodes)},\\
    g_n(L^{(i)}_{(N-n/2^{i+1})},L_N^{(i)},\hat{u}_{i}), &\text{($g_n$ nodes),}
  \end{cases}
\label{eq:08}
\end{equation}
where $f_n$ and $g_n$ functions were defined in \cite{Arikan} as:
\begin{equation}
  f_n(a,b)=\frac{1+ab}{a+b},
\label{eq:09}
\end{equation}
\begin{equation}
  g_n(a,b,\hat{u}_{i})=a^{1-2\hat{u}_{i}}b,
\label{eq:10}
\end{equation}
where $\hat{u}_{i}$ are the previous decoded bits. The  estimated
value $\Hat{u}_{i}$ is given by \eqref{eq:06}. Therefore, the
decision $g_n$ nodes depends on the estimate of $f_n$ nodes given by
\eqref{eq:09}, that is, of previously decoded bits. Other decoding
algorithms such as message passing techniques can be also considered
\cite{Chung,spa,jidf,mfsic,mbdf,bfidd,vfap,1bitidd,kaids,baplnc}.

\section{Gaussian Approximation}

The set $\mathcal{A}^\text{c}$ is obtained by the PC construction. The construction depends on several parameters, the main ones being: the length of the codeword $N$, the number of the information bits $K$, which channel will be used for transmission, the signal-to-noise ratio (SNR) target, or design-SNR, and the decoding approach. All construction methods covered in this paper will consider the AWGN channel and the SC decoder, mainly due to the large number of articles with results of PC construction with GA and SC decoding for AWGN channel, which allows for comparisons.

In the GA construction, the log likelihood ratio , \[ {\rm{ LLR}}(u_i)=log\left(\frac{W^{(i)}_N({\mathbf y}^N_1,\hat{u}_1^{i-1}|0)}{W^{(i)}_N({\mathbf y}^N_1,\hat{u}_1^{i-1}|1)}\right), \] is used as a Gaussian distribution function with a mean equal to half of the variance. Therefore, the mean of the LLRs is a sufficient statistic for their iterative update. The GA \cite{Chung} is given by
\begin{equation}
E\left(L^{(2i-1)}_N\right) = \phi^{-1}\left(1-\left(1-\phi\left(E\left(L^{(i)}_{N/2}\right)\right)\right)^2\right), \\
\label{eq:11}
\end{equation}
\begin{equation}
E\left(L^{(2i)}_N\right) = 2E\left(L^{(i)}_{N/2}\right),
\label{eq:12}
\end{equation}
with
\begin{equation}
L^{(0)}_1 = \frac{2}{\sigma^2}. 
\label{eq:13}
\end{equation}

The quantity $L_N^{(i)}$ denotes the LLR of the channel $W_N^{(i)}$, $\sigma^2$ and $E[\cdot]$ are the variance and the mean, respectively. In practice, in order to construct PC, we have $E[L_N^{(i)}]=L_N^{(i)}$. The function $\phi(x)$ is defined as:
\begin{equation}
\phi(x) =
\begin{cases}
1-\frac{1}{\sqrt{4\pi x}}\int \limits_{\mathbb{R}} {\rm tanh}\left(\frac{u}{2}\right)e^{\frac{-(x-u)^2}{4x}} \mathrm{d}u, & \ \text{if} \ x > 0,
\\
1, & \ \text{if} \ x = 0, \\
\end{cases}
\label{eq:14}
\end{equation}
where due to the integral function, we call it an Exact Gaussian Approximation (EGA) \cite{Fang}.

However, we have a complex integral function. The computational complexity will inevitably increase as the code length and polarization level increase. Moreover, the GA approach above has a numerical computation problem. The function $\phi(x)$ can arbitrarily approach zero as $x$ becomes very large. For example \cite{Ochiai}, for $x$ around 1000, a possible value in long code length construction, $\phi(x)$ can assume values lower than $10^{-100}$. We can solve the function $\phi(x)$ and  $\phi^{-1}(x)$ with the bisection method \cite{Oliveira5}. However, as $x$ becomes large, $\phi(x)$ becomes very small, which generates numerical inaccuracy, and consequently, generates an error in the code construction.

The author in \cite{Chung} also proposed a simplification of $\phi(x)$ by a two-segment approximation function described by
\begin{equation}
\phi(x)_{AGA} \approx
\begin{cases}
e^{-0.4527x^{(0.86)}+0.0218}, & \ \text{if} \ 0 < x \leq 10, \\
\sqrt{\frac{\pi}{x}}\left(1-\frac{10}{7x}\right)e^{-\frac{x}{4}}, & \ \text{if} \ x > 10, \\
\end{cases}
\label{eq:15}
\end{equation}
which is the so-called AGA \cite{Fang}. For codes with long block lengths, AGA induces performance losses due to the approximation error caused by the difference between $\phi(x)$ and $\phi(x)_{AGA}$ for $x=0$, that is, the AGA approach above has a numerical computation problem.
\begin{equation}
\phi(0)_{AGA}=e^{0.0218} > \phi(0)=1,
\label{eq:16}
\end{equation}
Equation \eqref{eq:16} shows the approximation error of $\phi(0)_{AGA}$.

A detailed analysis of the approximation error of equation \eqref{eq:15} and its effects on large block lengths can be found in \cite{Dai} and \cite{Ochiai}. Additionally, the AGA algorithm implements the calculation of transcendental, inverse and complex recursive functions, which can be avoided.

The computational complexity of the EGA construction is given by $\mathcal{O}(Nm)$, where $N$ is the length of the code and $m$ is the number of iterations to calculate the numerical solution of the integration and the function inversion, both from \eqref{eq:11}. The larger the value of $m$, the more accurate the numerical solution of integration and function inversion will be. The proposed approximations reduces the computational complexity of constructing the polar code to $\mathcal{O}(N)$, similar to the computational complexity in \cite{Arikan}. Note that the cost is associated to the design phase of PC. Once the codes are designed the operation cost is the same for all designs.

\section{Piecewise Gaussian Approximation}

It is known that EGA was originally proposed to design LDPC codes \cite{Chung} and when applied to the construction of PC \cite{Trifonov} it generates codes with good performance. However, it was not known if EGA \eqref{eq:14} can be improved for the construction of PC. From \eqref{eq:14}, for the purpose of analysis we define the function:
\begin{equation}
\psi(x,u) = \text{tanh}\left(\frac{u}{2}\right) \frac{1}{\sqrt{4 \pi x}} e^{\frac{-(u-x)^2}{4x}},
\label{eq:17}
\end{equation}
where we notice that the compound function is the product of a Gaussian function ($g(x,u)=\frac{1}{\sqrt{4 \pi x}} e^{\frac{-(u-x)^2}{4x}}$) with a hyperbolic tangent function (tanh$(\frac{u}{2})$), which we will call the Modified Gaussian (MG) function. An example of tanh$(\frac{u}{2})$ and $g(x,u)$ is shown in Fig. 3, for $x \approx 0.08$.

\begin{figure}[htb]
\begin{center}
\includegraphics[scale=0.55]{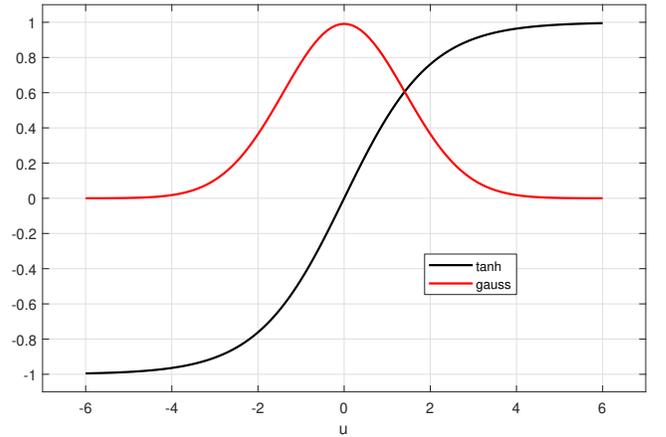}
\caption{Example of tanh and Gaussian function.}
\end{center}
\label{figura:fig02}
\end{figure}

The tanh$(\frac{u}{2})$ is an odd and zero-centered function, i.e., tanh$(\frac{u}{2})\subset[-1,1]$, with tanh$(\frac{u}{2})$ approaches -1 for the interval $\frac{u}{2} \in (-\inf,-4)$ and tanh$(\frac{u}{2})$ approaches +1 for the interval $\frac{u}{2} \in (+4,\inf)$. In Fig. 4 we can notice the compound function $\psi(x,u)$ in \eqref{eq:17}, and its behavior for several values of $x$ (mean), varying $u$, since tanh$(\frac{u}{2})$ does not depend on $x$, only on the factor $u$ as noted in equation \eqref{eq:17}. Each curve represents a $g(x,u)$ function with $x$ ranging from 0 to 14. Note that tanh$(\frac{u}{2})$ for $u < 1$ has greater importance than $g(x,u)$, for small values of $x$ and for $x$ approaching 0. Moreover, with the increase in $x$, the behavior tends to be of a $g(x,u)$ function, i.e, the Gaussian function becomes more dominant.

\begin{figure}[htb]
\includegraphics[scale=0.50]{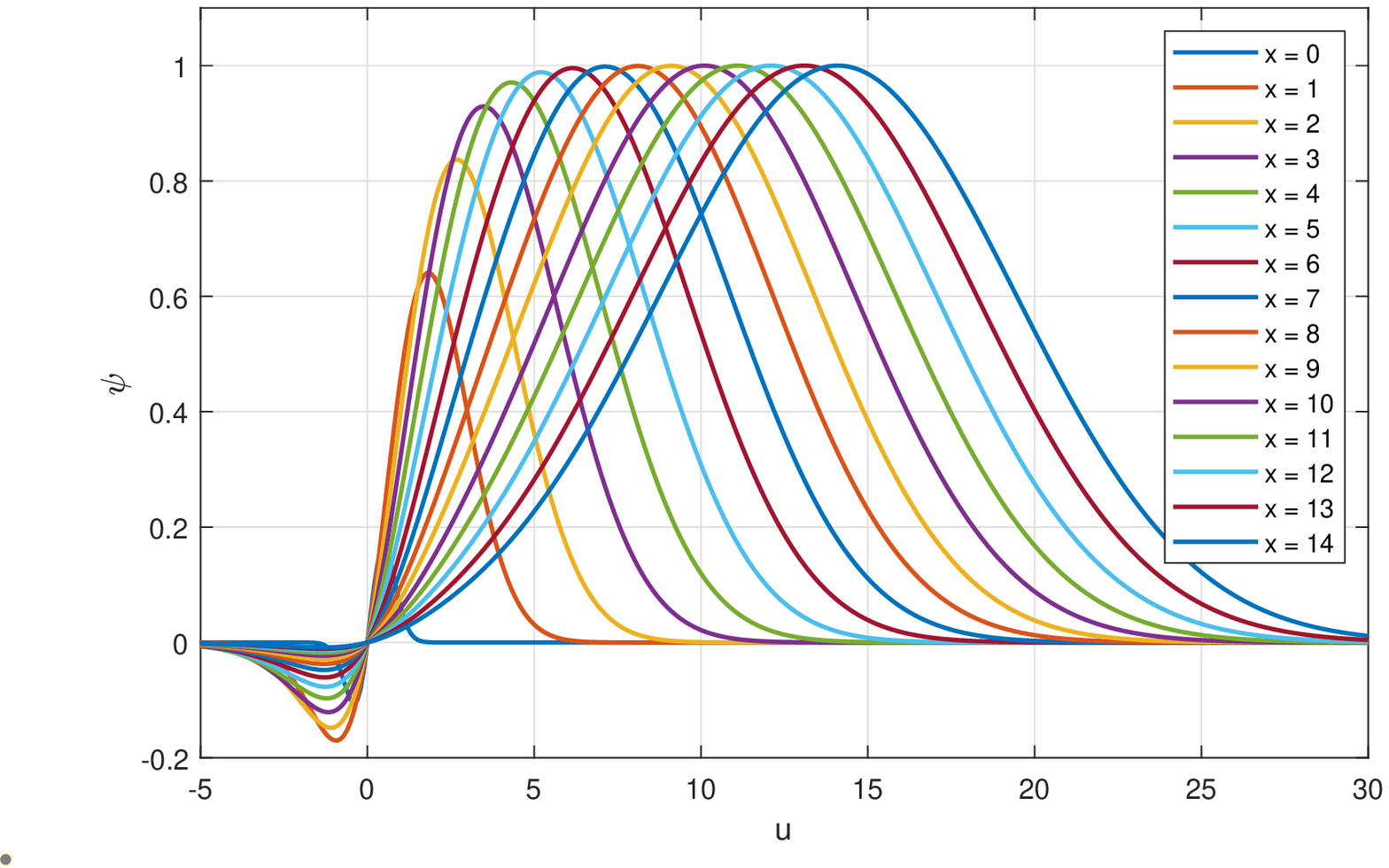}
\caption{Function $\psi$ in \eqref{eq:17}, with increment of x (mean) tends to the behavior of $g(x,u)$, ie, Gaussian.}
\label{figura:fig03}
\end{figure}

As explained in \cite{Chung}, to maintain the accuracy of EGA, it is important to preserve the symmetry condition \cite{Richardson}, expressed as $F(x)=F(-x)e^{-x}$, where $F(x)$ is the density of an LLR message. For the function $g(x,u)$ this condition can only be met by the mean. Observing $\psi(x,u)$, we have that the symmetry condition is preserved independently of the tanh$(\frac{u}{2})$ function.

Thus, to improve the performance of the GA construction for PC, we present a piecewise approximation for the function $\phi$ in \eqref{eq:14} and a new function is proposed to replace the function tanh$(\frac{ u}{2})$. In this approximation, we use an exponential function with the following terms: $a \cdot e^{b \cdot x}+c \cdot e^{d \cdot x}$.

The proposed piecewise ($\phi_p$) function optimized for PC is
\begin{equation}
\phi_p(x) =
\begin{cases}
1-\frac{1}{\sqrt{4\pi x}}\int \limits_{\mathbb{R}} f\left(\frac{u}{2}\right)e^{\frac{-(u-x)^2}{4x}} \mathrm{d}u, & \ x > 0, \\
1, & \ x = 0,
\end{cases}
\label{eq:18}
\end{equation}
with

\begin{equation}
f(x) =
\begin{cases}
 a \cdot e^{b \cdot x} + c \cdot e^{d \cdot x}, & \ x \geq -3.1 \ and \ x \leq 3.1, \\
+1, & \ x > +3.1, \\
-1, & \ x < -3.1.
\end{cases}
\label{eq:19}
\end{equation}

The parameters $a, b, c$, $d$ and cutoff at $x = \pm 3.1$ was obtained by exhaustive search for minimum FER in a PC(512,128), PC(1024,512) and PC(2048,1024) constructed with equations \eqref{eq:18} and \eqref{eq:19}, with design-SNR $\in (0,1,2,3)$, SNR $\in (0,1,2,3,4)$ and $500000$ iterations.
Initially, the odd function $f(x)$ is defined as $f(x)\subset[-1,1]$, $x \in \mathbb{R},$
$\lim_{x\to -\infty} f(x)=-1$,  $\lim_{x\to +\infty} f(x)=+1$ and $f(0)=0$, according to
\begin{equation}
\min_{\substack{ f(x) \\
}} (\text{FER}).
\label{eq:20}
\end{equation}

Generalization is possible because the function obtained in \eqref{eq:19} has the same initial format as tanh, which is the starting point for the optimization. Parameters a, b, c and d are continuously adjusted until the best FER performance is obtained. During the exhaustive search, the function $f(x)$ was tested with various formats, some of these formats are known functions. Some of these functions are represented in Fig. 5.

\begin{figure}[h]
\begin{center}
\includegraphics[scale=0.5]{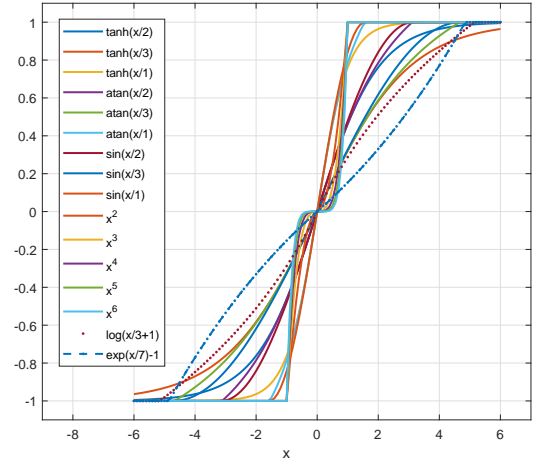}
\caption{Some $f(x)$ functions tested while exhaustive search in \eqref{eq:20}. For $x < 0$, consider $-x^2$, $-x^4$, $-x^6$, $-log(-x/3+1)$ and $exp(-x/7)-1$.}
\end{center}
\label{figura:fig04}
\end{figure}

The optimized parameters are $a = 1.9e+07$, $b = 8.4e-09$, $c = -1.8e+07$ and $d = -8.5e-09$.

In order to represent the PGA construction, we have updated equation \eqref{eq:11} as
\begin{equation}
E\left(L^{(2i-1)}_N\right) = \phi_p^{-1}\left(1-\left(1-\phi_p\left(E\left(L^{(i)}_{N/2}\right)\right)\right)^2\right), \\
\label{eq:21}
\end{equation}

Using the same format and limits proposed by \cite{Jeongseok}, using Root Mean Square Error (RMSE), we develop an approximation to the proposed function $\phi_p$ in \eqref{eq:18} given by
\begin{equation}
\phi_p(x) \approx
\begin{cases}
e^{-0.0484x^2-0.3258x}, & \ 0 \leq x < 0.867861, \\
e^{-0.4777x^{(0.8512)}+0.1094}, & \ 0.867861 \leq x < 10, \\
\sqrt{\frac{\pi}{x}}\left(1-\frac{1.509}{x}\right)e^{-\frac{x}{3.936}}, & \ x \geq 10. \\
\end{cases}
\label{eq:22}
\end{equation}
The values of $x$ are chosen to improve the approximation given by equation \eqref{eq:15} around $x = 0$, where $\phi(0)_{AGA} > 1$, equation \eqref{eq:16}. Thus, this approximation improves accuracy, but it is still constituted by transcendental functions.

In Fig. 6, we can observe the functions in \eqref{eq:11} with the  $\phi$ for EGA in \eqref{eq:14}, and the the functions in \eqref{eq:21} with the $\phi_p$ for PGA in \eqref{eq:18}. In the next section we will see the performance improvement due to this difference between the EGA and PGA functions.

\begin{figure}[htb]
\begin{center}
\includegraphics[scale=0.6]{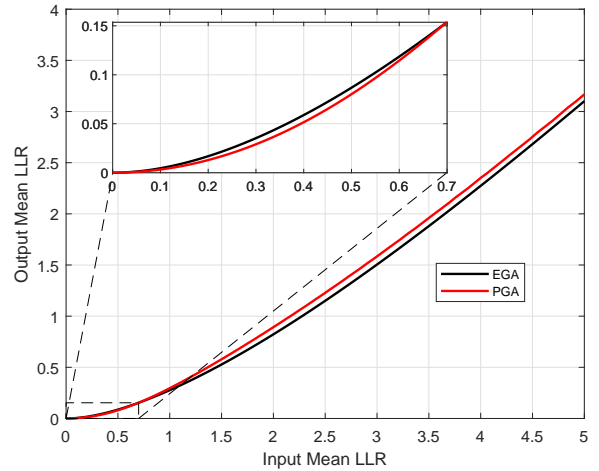}
\caption{Comparison of EGA and PGA with input mean LLR. }
\end{center}
\label{figura:fig05}
\end{figure}

\section{Proposed PGA Design Techniques}

In this section we present a detailed study of PGA by investigating the behavior of equation $\psi(x,u)$ in \eqref{eq:17}. We identify the key points and the statistical distribution of the results. Based on the statistics and in order to approximate PGA, we eliminate the transcendental functions and the inverse function. The approximation is generated with polynomial functions and we propose APGA for PC construction. Using a similar strategy, we propose an approximation for long blocks, called SPGA. We remark that APGA is an approximation of PGA, which uses the function $\phi_p(x)$ in \eqref{eq:18} and is optimized for medium blocks, whereas SPGA is an approximation of EGA, which uses the functions $\phi(x)$ in \eqref{eq:14} and has been optimized for long blocks.

\subsection{Modified Gaussian analysis}

In a more detailed analysis of the $\psi(x,u)$  function in \eqref{eq:17}, we can notice in Fig. 7b that from $x > 20$, that is, for the mean greater than 20, the behavior of $\psi(x,u)$ is Gaussian, i.e., \[ \{x>20 \ \text{and} \ u \in (-\inf,+\inf) \ | \ \psi(x,u) \approx g(x,u) \}. \] It means that from that point on the equation $\psi(x,u)$ can be well approximated by a polynomial function of degree 1, i.e., \[\{x>20 \ | \ \int \limits_{\mathbb{R}} \psi(x,u)du \approx 1 \}.\]

Next, we observe that in $\psi(x,u)$ from $x \in [6,20]$, the maximum point is identical to that of $g(x,u)$, as reproduced in Fig. 7a, i.e, $\{x \in [6,20] \ \text{and} \ u \in (-\inf,+\inf) \ | \ max(\psi(x,u)) = max(g(x,u)) \}$. Here we have one more key point at 6 and an important interval of study for the mean between 6 and 20.

\begin{figure}[htb]
\begin{center}
\includegraphics[scale=0.6]{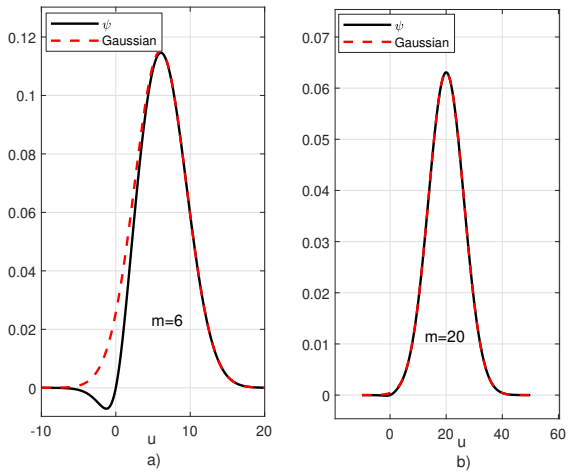}
\caption{We can observe that from $x=6$ the equations $\phi(x,u)$ and $g(x,u)$ already have the same maximum value, as observed in a), and are completely equal in b), for $x=20$.}
\end{center}
\label{figura:fig06}
\end{figure}
As observed in Fig. 4, for $x < 1$ the tanh$(\frac{u}{2})$ is more important than $g(x,u)$. This behavior can be better seen in Fig. 8, where for each figure the scale was reduced by $10^{-5}$.

\begin{figure}[htb]
\begin{center}
\includegraphics[scale=0.6]{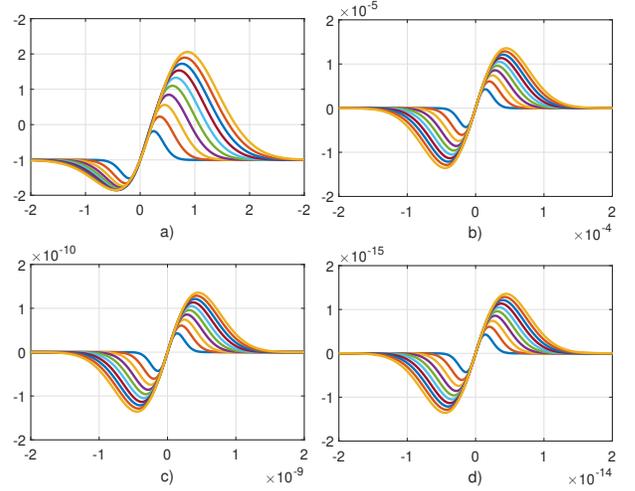}
\caption{Equation $\psi$ \eqref{eq:17} behavior around zero, for each sub-figure the scale was reduced by $10^{-5}$.}
\end{center}
\label{figura:fig07}
\end{figure}

According to our analysis, we observe that points 1, 6 and 20 are of fundamental importance for understanding the behavior of the $\phi$ function, for they mark the points at which $\phi$ approaches until it equals $g(x,u)$.

\subsection{Statistical analysis of the PGA function.}

Once the key points in the previous section are obtained, we need to perform a statistical analysis of PGA, investigating the distribution of the LLRs obtained by equation \eqref{eq:21}. For the design of the simplified approximation, we must make sure that we will have a good accuracy in the regions with the highest concentration of LLRs.

Let us then examine the statistical concentration of LLRs, considering intervals that include points 1, 6 and 20 as limits. See in Fig. 9 all LLRs generated by PGA for PC with lengths $N=256, 512,1024, 2048$. Note that there is a concentration in the range $[0,1]$. Looking in more detail the concentration of LLRs in the interval $[0,1]$, it is observed that there is always a greater concentration around zero for LLRs $\to 0$. This is because for $LLR < 1$, the PGA generates new LLRs closer and closer to zero because it uses a squared term in \eqref{eq:21}. This implies a greater resolution of the approximation in this interval.
\begin{figure}[htb]
\begin{center}
\includegraphics[scale=0.6]{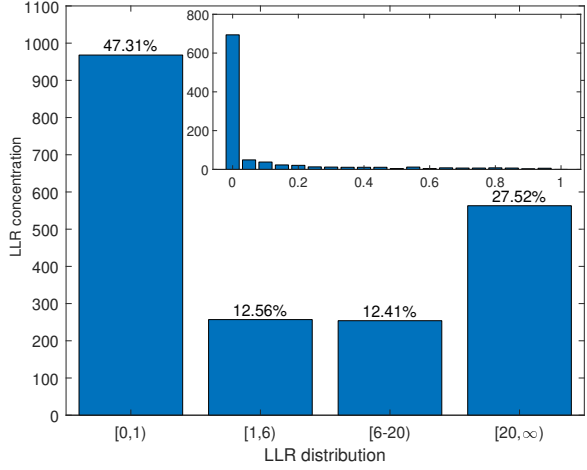}
\caption{LLR statistical distribution for PGA construction.}
\end{center}
\label{figura:fig08}
\end{figure}

As a result of this analysis, we suggest one more key point, with a value of 0.2, to be used in the intervals for the simplified approximation.

\subsection{Approximate PGA for medium blocks}

With the parameter obtained from the analysis of the behavior of the function $\psi$ and the statistical distribution of the LLRs of PGA, we propose to approximate the PGA in \eqref{eq:21} with a piecewise polynomials form given by

\begin{equation}
E[L_N^{(2i-1)}]=A(E[L^{(i)}_{N/2}]),
\label{eq:23}
\end{equation}
\begin{equation}
E[L_N^{(2i)}]=2E[L^{(i)}_{N/2}],
\label{eq:24}
\end{equation}
with

$A(x) =$
\begin{equation}
    \begin{cases}
      0.323x^2, & \ x \leqslant 0.2, \\
      -0.1x^3+0.43x^2-0.039x-0.005, & \ 0.2 < x \leqslant 1, \\
      -0.003x^3+0.063x^2+0.432x-0.2, & \ 1 < x \leqslant 6, \\
      -0.0002x^3+0.012x^2+0.777x-1.023, & \ 6 < x \leqslant 20, \\
      0.9803x - 2.109, & \  x > 20,
    \end{cases}
\label{eq:25}
\end{equation}
which is denoted as APGA. This approximation was obtained by minimum squared error curve-fitting. This operation involves only summations and multiplications, and avoids any transcendental functions. In Fig. 10a we can notice the accuracy of APGA in relation to PGA.

\subsection{Simplified PGA for long blocks}

Using the same approximation strategy and analysis as in the previous subsection, we propose the SPGA design technique for polar codes with long blocks. Similar to the approximations obtained previously, our objective is to obtain a simplified polynomial multi-segment function, using for this the same limits of equation \eqref{eq:25}. The proposed function has been designed by minimum squared error curve-fitting, with only addition and multiplication operations, without any transcendental functions. The simplified approximation is given by

\begin{equation}
E[L_N^{(2i-1)}]=S(E[L^{(i)}_{N/2}]),
\label{eq:26}
\end{equation}
\begin{equation}
E[L_N^{(2i)}]=2E[L^{(i)}_{N/2}],
\label{eq:27}
\end{equation}
with

$S(x) =$
\begin{equation}
    \begin{cases}
      -0.256x^3+0.461x^2+0.002x, & \ x \leqslant 0.2, \\
      -0.064x^3+0.294x^2+0.05x-0.004, & \ 0.2 < x \leqslant 1, \\
      -0.005x^3+0.092x^2+0.316x-0.133, & \ 1 < x \leqslant 6, \\
      0.002x^2+0.908x-1.588, & \ 6 < x \leqslant 20, \\
      0.995x - 2.459, & \ x > 20,
    \end{cases}
\label{eq:28}
\end{equation}
which is denoted as SPGA.

The EGA and PGA functions are distinct functions, as shown in Fig. 6, and two approximation functions, APGA for PGA and SPGA for EGA, are shown in Fig. 10a and Fig. 10b, respectively.

\begin{figure}[htb]
\begin{center}
\includegraphics[scale=0.6]{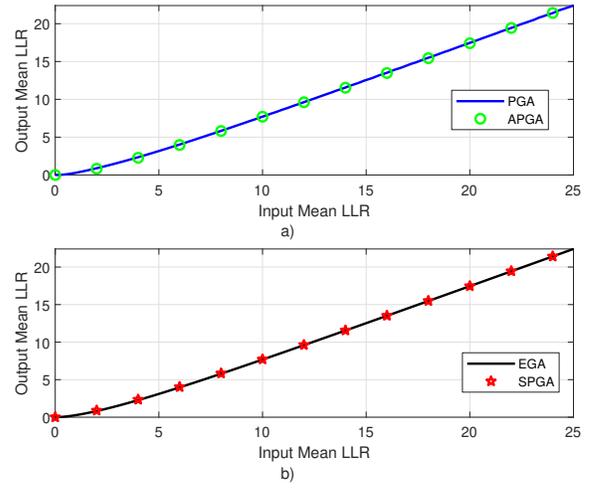}
\vspace{-1 em}
\caption{a) Accuracy of the APGA approximation to the PGA, and b) Accuracy of the
SPGA approximation to the EGA.}
\end{center}
\label{figura:fig09}
\end{figure}

\subsection{General algorithm for APGA and SPGA}

Here, we present a general construction algorithm that can be used for both APGA and SPGA. In particular, note that lines 11 and 12 of Algorithm 1 represent, respectively, the use of equations \eqref{eq:25} and \eqref{eq:28}. We can observe the mathematical simplification obtained when compared with equations \eqref{eq:11}, \eqref{eq:15} and \eqref{eq:22}, that is, without the need to calculate an inverse function and without transcendental functions.

\begin{algorithm}
\caption{General algorithm for APGA and SPGA}
\begin{algorithmic}[1]
\State $\text{Input}$: $N$, code length
\State $\text{Input}$: $K$, information bits
\State $\text{Input}$: design-SNR $E_{dB}=(RE_b/N_o)$ in dB
\State $\text{Output}$: $\text{F} = \{0,1,\ldots,N-1\}$ with
$|\text{F}|=N$
\State $S=10^{EdB/10}$
\State $n=log_2N$
\State $W \in \mathbb{R}^N$$, W(0)=4S$
\For{i = 1 to $n$}
   \State $d=2^i$
   \For{j = 1 to $\frac{d}{2}-1$}
       \State $W(d/2+\text{j}) = A(W(\text{j}-1)); \ {\rm {for}} \ {\rm {APGA}} \ \eqref{eq:25} \ {\rm {or}}$
       \State $W(d/2+\text{j}) = S(W(\text{j}-1));\ {\rm {for}} \ {\rm {SPGA}} \ \eqref{eq:28}$
       \State $W(\text{j}) = 2W(\text{j}-1)$
    \EndFor $\text{end for}$
\EndFor $\text{end for}$
\State F \ = \ Sorts \ W  \ indices \  in  \ ascending \ order \
\end{algorithmic}
\end{algorithm}

Recall that the objective of the proposed APGA and SPGA methods is to recursively calculate the reliability of each channel and all of them can be implemented with just a few calculations, which may involve non-linear functions. The overall complexity is proportional to the length of the codeword $N$.

\subsection{Accumulative Design Error}

The number of different positions (NDP) \cite{Kern} is a measure of the dispersion in PC construction. We used NDP to evaluate the approximation accuracy when comparing APGA and PGA; and also to evaluate the approximation accuracy when comparing SPGA to EGA. According to \cite{Kern}, we define as reference the set of frozen bits of PGA and EGA, called the reference set of frozen bits ($\mathcal {A}_c^{ref}$). We use it as a reference to compare the sets of frozen bits ($\mathcal{A}_c$) of all the construction methods. The number of different positions between $\mathcal{A}_c$ and $\mathcal{A}_c^{ref}$ is defined by
\begin{equation}
|\mathcal{A}_c \setminus \mathcal{A}_c^{ref}|  \triangleq  |\{x \in \mathcal{A}_c : x \notin \mathcal{A}_c^{ref}\}|.
\label{eq:29}
\end{equation}

We can use NDP as an indication of the quality of the approximation for given $n$. The smaller the NDP measure, the smaller the number of different frozen positions between $\mathcal{A}_c$ and $\mathcal{A}_c^{ref}$, which can lead to better FER for $\mathcal{A}_c$, closer to the FER with the ideal positions $\mathcal{A}_c^{ref}$. This approach is more effective for measuring the quality of approximations because we effectively compare the PC construction design.

We then define a mathematical expression for the channel difference with a unified index to compare the approximation methods for long blocks, which we call Accumulative Design Error (ADE) and whose main property is to account for the NDP. We define ADE as
\begin{equation}
\text{ADE}(n) =  \sum_{i=1}^{n} X^{i},
\label{eq:30}
\end{equation}
where $n = \log_2N$ and $X$ is the NDP for $n$. We have that the set $\mathcal{A}_c$ has different values for each approximation method. We can say that there is an optimal polynomial multi-segment approximation of \eqref{eq:11}, such that \[ \lim_{n \to \infty}\text{ADE}(n)=0, \] but the computational cost is prohibitive. However, we can consider that there is a two sub-optimal polynomial  approximation, which is feasible and has low computational cost, called $p_a$ and $p_b$, where we get $\mathcal{A}_c^a$ and $\mathcal{A}_c^b$, respectively. Since $\exists$ $n>1$ such that
\begin{align}
\sum_{i=1}^{n} |\mathcal{A}_c^a \setminus \mathcal{A}_c^{ref}|(i) &< \sum_{i=1}^{n} |\mathcal{A}_c^b \setminus \mathcal{A}_c^{ref}|(i), \ {\rm or} \nonumber \\
\sum_{i=1}^{n} \text{NDP}_a(i) &< \sum_{i=1}^{n} \text{NDP}_b(i), \ {\rm or} \nonumber  \\
\sum_{i=1}^{n} X_a^{i} &< \sum_{i=1}^{n} X_b^{i},
\label{eq:31}
\end{align}
to $p_a$ being the most accurate approximation of equation in \eqref{eq:11}. The proof is given in Appendix A.

\section{Numerical results}

In this section, we evaluate the proposed APGA and SPGA construction techniques and compare them against existing approaches for several scenarios. In particular, we assess the performance of the proposed construction techniques for medium and long blocks.
We follow the terminology in terms of block lengths adopted by related work on PC for the code construction scenario, as noted in \cite{Fang}, \cite{Dai} and \cite{Ochiai}. In subsection A, “Designs for medium blocks”, the comparison is made for block lengths up to 2048 bits, as can be seen in Table 1 and in Fig. 11, 12 and 13. In subsection B, “Designs for long blocks”, we compare the methods for blocks with lengths greater than 4096 bits. In Tables 2 and 3, and NPD comparisons are made for blocks from 2048 bits to 131072 bits. Additionally, in Fig. 14 we compare the FER performance for blocks for $n=12$, $n=14$ and $n=16$, which is equivalent to long blocks with length of 4096 bits, 16384 bits and 65536 bits, respectively.
In the following, we illustrate the results of MC simulations, with the AFF3CT toolbox \cite{Cassagne}. We simulated for Binary Phase shift keying (BPSK), AWGN, SC and 1dB design-SNR modulation. The simulation loops adopt as stopping criterion the counting of $200$ frame errors. The exact GA can be calculated with extremely high accuracy through careful numerical integration.

\subsection{Designs for medium blocks}

In Table 1 we have the NDP between APGA and PGA, for $R =(1/2,1/3,2/3)$. The observed values are due to the application of APGA in the construction of PC. Note that NDP has an increasing trend as $N$ increases.

\begin{table}[htb]
\centering
\caption{NDP between APGA and PGA}
\begin{tabular}{|c|c|c|c|c|c|}
\hline
$R$ & 128  & 256  & 512  & 1024 & 2048 \\ \hline\hline
1/2 & 0    & 0    & 0    & 2    & 8    \\ \hline
1/3 & 0    & 0    & 0    & 2    & 6    \\ \hline
2/3 & 0    & 0    & 2    & 2    & 6    \\ \hline
\end{tabular}
\end{table}

In Fig. 11 we have the performance between PGA and EGA. For $N \geq 128$, the FER graph shows an increasing PGA gain with increasing $N$, being $0.25$ dB for $N = 2048$.

\begin{figure}[htb]
\begin{center}
\includegraphics[scale=0.6]{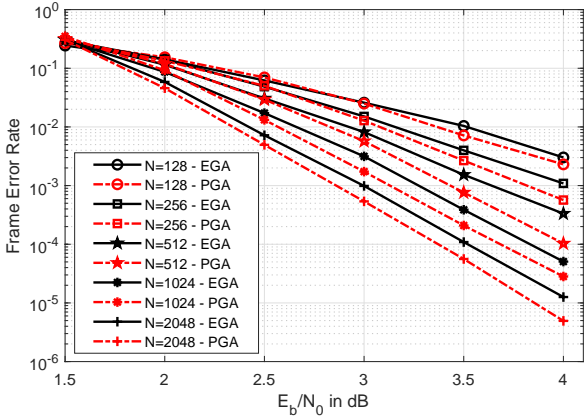}
\caption{FER performance between EGA and PGA, N=$2^n$, $n=(7,8,9,10,11)$ and $R=1/2$.}
\end{center}
\label{figura:fig10}
\end{figure}

The performance of PGA and EGA is shown in Fig. 12 for various block lengths. We remark the FER gain for $N \geq 256$, reaching $0.15$dB for $N=2048$. Note also that at $N=128$ there is no difference in PGA and EGA performance.

\begin{figure}[htb]
\begin{center}
\includegraphics[scale=0.6]{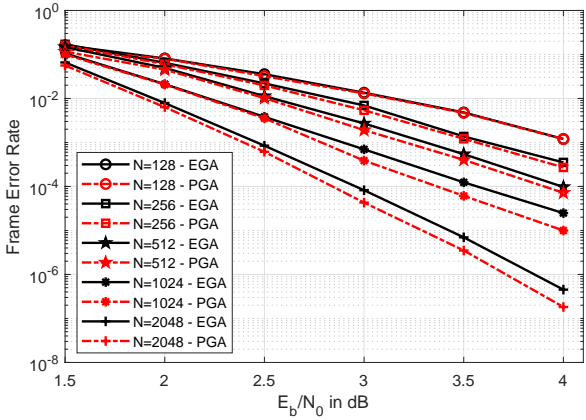}
\caption{FER performance between EGA and PGA, N=$2^n$, $n=(7,8,9,10,11)$ and $R=1/3$.}
\end{center}
\label{figura:fig11}
\end{figure}

Now, we compare PGA and EGA, with $R=2/3$ and various block lengths in Fig. 13. Note that there is a gain in terms of FER for $N \geq 512$ obtained by PGA, which reaches $0.25$dB for $N= 2048$.

Note that in Fig. 12 the FER for PGA and EGA at $N=128$ is the same, that is, for PC with $N=128$ and $R=1/3$ the set $\mathcal{A}$ obtained by the PGA method is the same set $\mathcal{A}$ obtained by the EGA method. The same can be seen in Fig. 13 for $N=128$ and for $N=256$, both with $R=2/3$, the PGA for $N=128$ is the same as EGA for $N=128$ and the PGA for $N=256$ is the same as EGA for $N=256$.

\begin{figure}[htb]
\begin{center}
\includegraphics[scale=0.6]{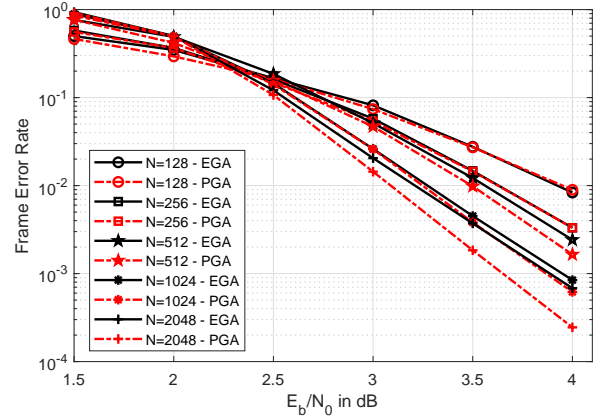}
\caption{FER performance between EGA and PGA, N=$2^n$, $n=(7,8,9,10,11)$ and $R=2/3$.}
\end{center}
\label{figura:fig12}
\end{figure}

We can observe that in the scenario of medium blocks, with PGA we observe a continuous improvement in the performance of FER when compared to EGA.

\subsection{Designs for Long Blocks}

An important aspect to be considered in the simulation for long blocks is the difficulty to observe in figures the difference in FER performance. Performance comparison by NDP can be used as a construction quality parameter. However, we observe that small NDP, that is, very small percentage differences of construction, are hardly observable in terms of FER curves, despite representing a better approximation. Another important aspect in comparing the approximation methods is the Root Mean Square Error (RMSE). It can be used effectively when the intervals used for approximation are equal, which is not the case in our analysis.

We present in the tables below the NDP difference between the construction of the SPGA and the constructions proposed by Trifonov (RGA) \cite{Trifonov2}, Fang (SGA) \cite{Fang}, Dai (AGA-4) \cite{Dai} and Ochiai (LGA) \cite{Ochiai}; for $R = 1/2, 1/3, 2/3$. We included in the comparison the approximation RGA due to its polynomial format, which meets the simplification requirements objective of this study. The approximation SGA follows the same strategy. This difference in channels, i.e., NDP, represents the quality of constructions in relation to the EGA. The most accurate construction achieves the smallest NDP, i.e, minus the difference for the optimal set of frozen bits ($\mathcal{A}_c^{ref}$). Initially, in Table 2, we have NDP for $R=1/2$. Observe that SPGA has lower NDP with the increase of $N$. This means that the SPGA approximation is more accurate than the others in this scenario.

\begin{table}[htb]
\centering
\caption{NDP for EGA with $R$ = 1/2}
\begin{tabular}{|c|c|c|c|c|c|ll}
\hline
$N$      & RGA                  & SGA             & AGA-4            & LGA               & SPGA  \\ \hline\hline 
2048   & 2                    & 4               & 2                & 2                 & 0     \\ \hline 
4096   & 6                    & 10              & 2                & 2                 & 0     \\ \hline 
8192   & 16                   & 18              & 2                & 2                 & 4     \\ \hline 
16384  & 30                   & 46              & 10               & 8                 & 10    \\ \hline 
32768  & 68                   & 90              & 12               & 16                & 16    \\ \hline 
65536  & 158                  & 176             & 42               & 42                & 34    \\ \hline 
131072 & 342                  & 332             & 78               & 84                & 60    \\ \hline 
\end{tabular}
\end{table}

In Table 3, for $R=2/3$, the fairest comparison is between SPGA, RGA and SGA, which are polynomial approximations. Among them, the SPGA remains with a lower NDP, that is, more accurate than the others. In turn, the design techniques AGA-4 and LGA have the smallest NDP, but it should be remarked that they are approximations of greater complexity which include transcendental functions.

\begin{table}[htb]
\centering
\caption{NDP for EGA with $R$ = 2/3}
\begin{tabular}{|c|c|c|c|c|c|}
\hline
$N$    & RGA              & SGA         & AGA-4      & LGA            & SPGA  \\ \hline\hline 
2048   & 6                & 0           & 0          & 0              & 0     \\ \hline 
4096   & 24               & 6           & 4          & 4              & 2     \\ \hline 
8192   & 54               & 18          & 4          & 4              & 6     \\ \hline 
16384  & 132              & 32          & 8          & 8              & 10    \\ \hline 
32768  & 386              & 96          & 22         & 24             & 28    \\ \hline 
65536  & 766              & 396         & 34         & 32             & 106   \\ \hline 
131072 & 1840             & 428         & 78         & 80             & 336   \\ \hline 
\end{tabular}
\end{table}

For $R=1/3$, as can be seen in Table 4, SPGA remains with the lowest NDP, which suggests that it is the most accurate approximation in this scenario. We can see the SPGA has the lowest NDP in the three scenarios, that is, for $R = 1/2, 1/3, 2/3$, and this characteristic is maintained with the increment of $N$. We can conclude that it is the most accurate approximation of EGA.

In fact, we can prove that the proposed approximations are better as evidenced by the ADE, which also indicates better performance. For example, for EGA in Table 4 we have NPD for EGA of the methods RGA, SGA, AGA-4, LGA and SPGA. According to equation (30), the ADE account for the NDP for each $n$. In other words, comparing with EGA, for each method, the difference of channels in the code construction is calculated for each $n$, and then we account for these differences. In this way, we can compare all construction methods in relation to EGA, and the smaller the ADE, the better the approximation for EGA. So, we have
\begin{align}
\text{ADE}_{\text{SPGA}}(17) &=  \sum_{n=1}^{17} X^{n}, \nonumber \\
&= X^{11}+X^{12}+X^{13}+X^{14}, \nonumber \\
&+X^{15}+X^{16}+X^{17}, \nonumber \\
&=0+4+0+10+14+32+44, \nonumber \\
&=104. \nonumber
\end{align}

It can be noted that $\text{ADE}_{\text{SPGA}}(17)$ has the lowest value among the methods in all analyzed scenarios, so
\begin{equation}
\text{ADE}_{\text{SPGA}}(17) < \text{ADE}_{\text{SGA}}(17) < \text{ADE}_{\text{RGA}}(17).
\label{eq:32}
\end{equation}

\begin{table}[htb]
\centering
\caption{NDP for EGA with $R$ = 1/3}
\begin{center}
\begin{tabular}{|c|c|c|c|c|c|c|}
\hline
$N$    & RGA              & SGA         & AGA-4      & LGA            & SPGA  \\ \hline\hline 
2048   & 6                & 4           & 2          & 2              & 0     \\ \hline 
4096   & 4                & 2           & 2          & 4              & 4     \\ \hline 
8192   & 10               & 10          & 10         & 8              & 0     \\ \hline 
16384  & 26               & 14          & 8          & 10             & 10    \\ \hline 
32768  & 48               & 26          & 18         & 12             & 14    \\ \hline 
65536  & 96               & 46          & 34         & 32             & 32    \\ \hline 
131072 & 192              & 118         & 72         & 66             & 44    \\ \hline 
\end{tabular}
\end{center}
\end{table}

In Fig. 14 we have the FER performance of the code construction alternatives observed in Table 4, for $R=1/3$. As the AGA-4 and LGA approximations have performance comparable to EGA, however they are more complex approximations with transcendental functions. We remark that the approximation AGA \cite{Chung} is not shown here because it would result in much worse performance than the others.

\begin{figure}[htb]
\begin{center}
\includegraphics[scale=0.6]{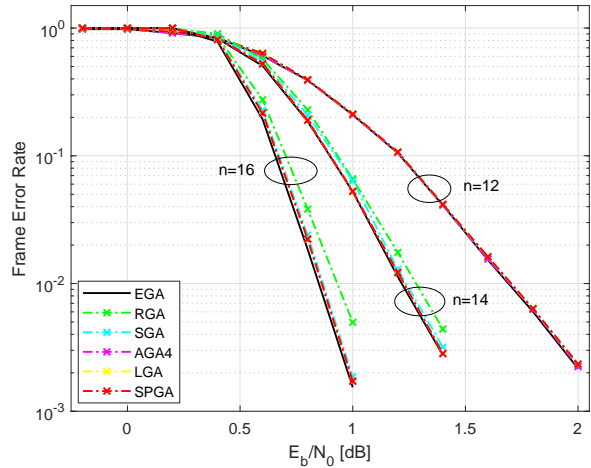}
\caption{FER comparison between EGA, RCA \cite{Trifonov2}, SGA \cite{Fang}, AGA-4 \cite{Dai}, LGA \cite{Ochiai} and SPGA, N=$2^n$, $n=(12,14,16)$ and $R=1/3$.}
\end{center}
\label{figura:fig13}
\end{figure}

It can be seen that the proposed SPGA method has better performance as compared to other code construction methods. This is because the SPGA method was optimized with the aid of the analysis of the $\psi$ equation in \eqref{eq:17}. All other methods work almost equally well, with less accurate approximations to EGA than SPGA. We can observe the similarity of performance between the approximations, despite the SPGA being the best approximation in the scenario considered.

\section{Conclusions}

In this work we have presented a novel method for polar code construction, called PGA, which is suitable for medium and long blocks. We have also presented APGA based on an analysis of the behavior of the PGA function, the identification of its key points and the analysis of the statistical distribution of their LLRs. Using the same analysis criterion, we have developed SPGA for long blocks, in the form of a multi-segment function. Moreover, we have introduced ADE as a figure of merit for comparison of designs because it effectively measures the difference in PC construction methods. The effectiveness of the APGA and SPGA approaches has been investigated by comparing them with other design techniques with the same order of complexity through simulations and analytical arguments using the ADE of the analyzed approximation techniques.

\begin{appendices}
\section{Proof equation 31}
According to Weierstrass Approximation Theorem \cite{Giardina}, suppose that $f:[a,b] \to \mathbb{R}$ is a continuous real-valued function defined on the real interval [a,b]. For every $\varepsilon > 0$, there exists a polynomial $p$ such that for all $x \in [a, b]$, we have

\begin{equation}
|f(x) - p(x)| < \varepsilon.
\nonumber
\end{equation}

Given the continuous and increasing function $f(x)$, being that $x_i < x_{i+1}$ implies $f(x_i) < f(x_{i+1})$, by simplifying the equation in \eqref{eq:11}, we have

\begin{equation}
f\left(x\right) = \phi^{-1}\left(1-\left(1-\phi\left(x\right)\right)^2\right).
\label{eq:33}
\end{equation}

Now, consider two polynomial approximations $p_a(x)$ and $p_b(x)$, such that
\begin{equation}
|f(x_i)-p_a(x_i)| < \varepsilon_a,
\label{eq:34}
\end{equation}
and
\begin{equation}
|f(x_i)-p_b(x_i)| < \varepsilon_b,
\label{eq:35}
\end{equation}
with $x_i \in [a, b]$ and approximation errors $\varepsilon_a$ and $\varepsilon_b$. Additionally, for the inequality below
\begin{equation}
|f(x_i)-p_a(x_n)| > \varepsilon_a,
\label{eq:36}
\end{equation}
for $\forall x_n \in [a, b]$, $\forall x_i \in [a, b]$ and $x_i \neq x_n$; and
\begin{equation}
|f(x_i)-p_b(x_n)| > \varepsilon_a,
\label{eq:37}
\end{equation}
for $\forall x_n \in [a, b]$ and $\forall x_i \in [a, b]$, including $x_i = x_n$, we have that $f(x)$ is strictly increasing, so we have the guarantee that
\begin{equation}
\varepsilon_a < \varepsilon_b.
\label{eq:38}
\end{equation}

This result can be generalized as $\forall p_j(x_n)$ and approximation errors $\varepsilon_j$, with $j \in \mathbb{N}$, so
\begin{equation}
|f(x_i)-p_j(x_n)| < \varepsilon_j,
\label{eq:39}
\end{equation}
and
\begin{equation}
|f(x_i)-p_j(x_n)| > \varepsilon_a, \ \text{for} \ \forall j,
\label{eq:40}
\end{equation}
then
\begin{equation}
\varepsilon_a < \varepsilon_j, \ \text{for} \ \forall j,
\label{eq:41}
\end{equation}
and if
\begin{equation}
|f(x_i)-p_{j}(x_n)| > \varepsilon_{j+1}, \ \text{for} \ \forall j,
\label{eq:42}
\end{equation}
then
\begin{equation}
\varepsilon_a < \varepsilon_j < \varepsilon_{j+1} < \dots < \varepsilon_{j+n}.
\label{eq:43}
\end{equation}

So, we have to give the sets
\begin{align}
\text{F} &= \{ f(x_1), f(x_2), \dots, f(x_n) \}, \nonumber \\
\text{P}_a &= \{p_a(x_1), p_a(x_2), \dots, p_a(x_n)\}, \nonumber \\
\text{P}_b &= \{ p_b(x_1), p_b(x_2), \dots, p_b(x_n) \}, \nonumber
\end{align}
with $x_i \in [a,b]$, $i \in [1,...,n]$ and conditions given in \eqref{eq:39}, \eqref{eq:40}, \eqref{eq:41} and \eqref{eq:42}, then we have to
\begin{align}
|f(x_i)-p_a(x_i)| &< \varepsilon_a < |f(x_i)-p_b(x_i)|, \nonumber\\
|f(x_i)-p_a(x_i)| &< \varepsilon_a < \varepsilon_b. \nonumber
\end{align}

Using these results, we have for the approximations RGA, SGA, SPGA:
\begin{align}
|f_{\text{EGA}}(x_i)-p_{\text{SPGA}}(x_n)| &< \varepsilon_1, \nonumber \\
|f_{\text{EGA}}(x_i)-p_{\text{RGA}}(x_n)| &< \varepsilon_2, \nonumber \\
|f_{\text{EGA}}(x_i)-p_{\text{SGA}}(x_n)| &< \varepsilon_3. \nonumber
\end{align}
and condition in \eqref{eq:42}
\begin{align}
|f_{\text{EGA}}(x_i)-p_{\text{SPGA}}(x_n)| &> \varepsilon_1, \nonumber \\
|f_{\text{EGA}}(x_i)-p_{\text{RGA}}(x_n)| &> \varepsilon_1, \nonumber \\
|f_{\text{EGA}}(x_i)-p_{\text{SGA}}(x_n)| &> \varepsilon_1. \nonumber \\
|f_{\text{EGA}}(x_i)-p_{\text{RGA}}(x_n)| &> \varepsilon_2, \nonumber \\
|f_{\text{EGA}}(x_i)-p_{\text{SGA}}(x_n)| &> \varepsilon_2. \nonumber
\end{align}
then, \eqref{eq:43},
\begin{equation}
\varepsilon_1 < \varepsilon_2 < \varepsilon_3
\label{eq:44}
\end{equation}

Equation \eqref{eq:44} determines the improved approximation of $p_{\text{SPGA}}$ over $p_{\text{RGA}}$ and $p_{\text{SGA}}$. We know that $\mathcal{A}_c^{ref}$ is obtained from $f_{\text{EGA}}$, the $\mathcal{A}_c^a$ is obtained from $p_{\text{SPGA}}$, the $\mathcal{A}_c^b$ is obtained from $p_{\text{RGA}}$ and the $\mathcal{A}_c^c$ is obtained from $p_{\text{SGA}}$. According to \eqref{eq:29} and the result in \eqref{eq:44} we can consider that

\begin{equation}
|\mathcal{A}_c^a \setminus \mathcal{A}_c^{ref}| < L,
\label{eq:45}
\end{equation}
\begin{equation}
|\mathcal{A}_c^b \setminus \mathcal{A}_c^{ref}| < L,
\label{eq:46}
\end{equation}
and
\begin{equation}
|\mathcal{A}_c^c \setminus \mathcal{A}_c^{ref}| < L,
\label{eq:47}
\end{equation}
with L $> 0$ and L $\in \mathbb{N}$ for a given $n$. Subtracting \eqref{eq:45} from \eqref{eq:46} and \eqref{eq:47}, we have
\begin{equation}
|\mathcal{A}_c^a \setminus \mathcal{A}_c^{ref}| < |\mathcal{A}_c^b \setminus \mathcal{A}_c^{ref}|  < |\mathcal{A}_c^c \setminus \mathcal{A}_c^{ref}|,
\label{eq:48}
\end{equation}
and for every set $n$, we have
\begin{equation}
\sum_{i=1}^{n} |\mathcal{A}_c^a \setminus \mathcal{A}_c^{ref}| < \sum_{i=1}^{n} |\mathcal{A}_c^b \setminus \mathcal{A}_c^{ref}| < \sum_{i=1}^{n} |\mathcal{A}_c^c \setminus \mathcal{A}_c^{ref}|.
\label{eq:49}
\end{equation}
Therefore, the above expression can be rewritten as equation \eqref{eq:31}, i.e.,
\begin{equation}
\sum_{i=1}^{n} X_a^{i} < \sum_{i=1}^{n} X_b^{i} < \sum_{i=1}^{n} X_c^{i}.
\label{eq:50}
\end{equation}

In fact, the inequality in \eqref{eq:50} will always hold under the conditions established in \eqref{eq:44}. Therefore, we can verify in Table 5 the RMSE among the polynomial approximation alternatives for the interval  $x \in$ [0,20].

\begin{table}[htb]
\centering
\caption{RMSE between RGA, SGA, SPGA and EGA}
\begin{tabular}{|c|c|c|c|}
\hline
RMSE & RGA   & SGA    & SPGA   \\ \hline
EGA  & 0,036 & 0,0338 & 0,0215 \\ \hline
\end{tabular}
\end{table}

We can observe that SPGA has the smallest RMSE which implies the best ADE and, therefore, the best performance.

\end{appendices}

\end{document}